\documentclass{aa}

\usepackage{graphics}

\begin{document}

\title{PSR B\,1706$-$44 and the SNR G\,343.1$-$2.3 as the 
remnants of a cavity supernova explosion}

\author{D.C.-J.\,Bock
\inst{1}
\and
V.V.\,Gvaramadze
\inst{2,3,4}\thanks{{\it Address for
correspondence}: Krasin str. 19, ap. 81, Moscow, 123056, Russia
(vgvaram@mx.iki.rssi.ru)}
\institute{Radio Astronomy Laboratory, University of California, 
Berkeley, CA 94720, USA
\and
Sternberg State Astronomical Institute, Moscow State
University, Universitetskij Pr.~13, Moscow, 119992,
Russia
\and
E.K.Kharadze Abastumani Astrophysical Observatory, Georgian Academy of
Sciences, A.Kazbegi ave.~2-a, Tbilisi, 380060, Georgia
\and
Abdus Salam International Centre for 
Theoretical Physics, Strada Costiera 11, P.O.\ Box 586, 34100 Trieste, Italy
}
}
\offprints{D.~C.-J.~Bock, \email{dbock@astro.berkeley.edu}} 

\date{Received 9 April 2002 / accepted 2 August 2002}

\titlerunning{PSR B\,1706$-$44 and the SNR G\,343.1$-$2.3}
\authorrunning{Bock \& Gvaramadze}

\abstract{ The possible association of the supernova remnant (SNR)
  \object{G\,343.1$-$2.3} with the pulsar \object{PSR B\,1706$-$44}
  (superposed on the arclike ``shell" of the SNR) has been questioned
  by some authors on the basis of an inconsistency between the implied
  and measured (scintillation) transverse velocities of the pulsar,
  the absence of any apparent interaction between the pulsar and the
  SNR's ``shell", and some other indirect arguments.  We suggest,
  however, that this association could be real if both objects are the
  remnants of a supernova (SN) which exploded within a mushroom-like
  cavity (created by the SN progenitor wind breaking out of the parent
  molecular cloud). This suggestion implies that the actual shape of
  the SNR's shell is similar to that of the well-known SNR
  \object{VRO\,42.05.01} and that the observed bright arc corresponds
  to the ``half" of the SNR located inside the cloud. We report the
  discovery in archival radio data of an extended ragged radio arc to
  the southeast of the bright arc which we interpret as the
  ``half'' of the SN blast wave expanding in the intercloud
    medium.
  \keywords{Stars: neutron -- pulsars: individual: PSR B\,1706$-$44 --
    ISM: bubbles -- ISM: individual objects: G\,343.1$-$2.3 -- ISM:
    supernova remnants} }

\maketitle

\section{Introduction}
%

The pulsar \object{PSR B\,1706$-$44} (Johnston et al.\ \cite{joh92})
is superposed on an incomplete arc of radio emission (McAdam et al.\ 
\cite{mca93}).  McAdam et al.~interpreted this arc as a shell-type
supernova remnant (SNR), named \object{G\,343.1$-$2.3}, and suggested
that the SNR is physically associated with PSR B\,1706$-$44. This
suggestion was questioned by Frail et al.\ (\cite{fra94a}) and
Nicastro et al.\ (\cite{nic96}; see, however, Dodson et al.\ 
\cite{dod01}).  Usually, a particular claimed pulsar/SNR association
is considered reliable if the following five criteria are fulfilled
(e.g. Kaspi \cite{kas96}):
\begin{enumerate}
\item agreement of independent distance estimates for pulsar and SNR;

\item agreement of independent age estimates for pulsar and SNR;

\item consistence of the implied pulsar transverse velocity (i.e.~the
  velocity inferred by the displacement of the pulsar from the
  geometrical centre of the associated SNR) with the measured (proper
  motion and/or scintillation) velocity;

\item existence of any sign of interaction between the pulsar and the SNR;

\item ``correct" (inferred or measured) orientation of the vector of
  pulsar transverse velocity (it is assumed that this vector should be
  pointed away from the geometrical centre of the associated SNR).
\end{enumerate}

Although the distance and age estimates for PSR B\,1706$-$44 and
G\,343.1$-$2.3 are in reasonable agreement, the implied transverse
velocity is at least an order of magnitude larger than the
scintillation velocity calculated by Nicastro et al.\ (\cite{nic96}).
This inconsistency along with the absence of any apparent interaction
between the pulsar and the SNR constitute the two main arguments
against the physical association between these two objects (Frail et
al.  \cite{fra94a}, Nicastro et al.\ \cite{nic96}). The fifth
criterion is not applied to the system, since the direction of the
pulsar proper motion is still unknown (cf. Giacani et al.\ 
\cite{gia01} with Frail et al.\ \cite{fra94a}; see also Sect.  3.5).
Additional (indirect) arguments against the association are based on
Gaensler \& Johnston's (\cite{gae95}) statistical study, which
suggests that young pulsars cannot overrun their parent SNR shells
(Nicastro et al.\ \cite{nic96}; see however Arzoumanian et
  al.\ \cite{arz02}) and on the large extent of the ``halo" around
the pulsar (Frail et al.\ \cite{fra94a}).

In this paper we show how the existing observational data on PSR
B\,1706$-$44 and G\,343.1$-$2.3 can be interpreted in favour of their
physical association (Sect.~2) and discuss the criteria for evaluating
the reliability of pulsar/SNR associations as applied to this system
(Sect.~3).  The main suggestion of the paper is that the association
between PSR B\,1706$-$44 and the SNR G\,343.1$-$2.3 could be real if
both objects are the remnants of a SN which exploded within a
mushroom-like cavity created by the SN progenitor wind breaking out of
the parent molecular cloud (Sect.~2.2).  This suggestion implies that
in addition to the known bright ``half" of the SNR G\,343.1$-$2.3
there should exist a more extended and weaker component, so that the
actual shape of G\,343.1$-$2.3 is similar to that of the well-known
SNR \object{VRO\,42.05.01}. It is remarkable that the 2.4 GHz Parkes
Survey of Duncan et al.\ (\cite{dun95}) shows the existence of such an
extended component.

\section{The SNR G\,343.1$-$2.3}

Before discussing the criteria for evaluating the reliability of the
association between PSR B\,1706$-$44 and the SNR G\,343.1$-$2.3 we
review the observational data on this system and propose a
scenario for its formation.

\subsection{Observational data}

SNR G\,343.1$-$2.3 was discovered by McAdam et al.\ (\cite{mca93}).
Their 843 MHz image of G\,343.1$-$2.3 shows a well-defined arc (a
half-ellipse) of radio emission with the brightest (northern) part
closest to the Galactic plane. The maximum extent of the arc is about
$40'$.  A VLA image of the SNR obtained by Frail et
al.~(\cite{fra94a}) shows the existence of weak, diffuse emission both
inside and outside the bright arc.  This emission fills a region
similar to and about two times more extended than the bright arc
(Dodson et al.\ \cite{dod01}; see also Duncan et al.\ \cite{dun95} and
Fig.~\ref{plot}).  The ROSAT observations of the field around PSR
B\,1706$-$44 do not reveal any sign of correlation between the soft
(0.1--2.4 keV) diffuse X-ray emission and the radio emission of the
SNR (Becker et al.\ \cite{bec95}).  There are no reported optical
observations of the SNR.  PSR B\,1706$-$44 is superposed on the
outside edge of the bright radio arc and there are no morphological
signatures of interaction between them.  The pulsar appears to be
surrounded by a radio nebula of about $3'$ in size (Giacani et al.\ 
\cite{gia01}; cf.~Frail et al.\ \cite{fra94a}).  Giacani et al.\ 
suggest that this nebula is powered by the pulsar, on the basis of the
nebula's flat radio spectral index ($\simeq 0.3$) and the high mean
fractional polarization ($\simeq 20\%$) of its radio emission.
 
\subsection{A scenario for the origin of system PSR B\,1706$-$44/G\,343.1$-$2.3}

\begin{figure*}
  \resizebox{12cm}{!}{\includegraphics{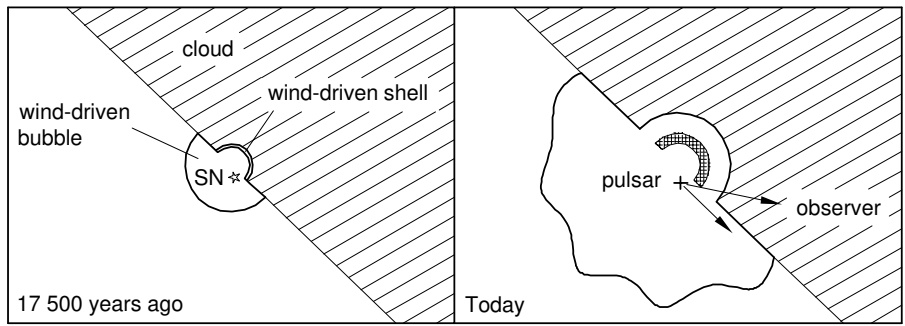}}
  \caption{Schematic of the proposed origin of G\,343.1$-$2.3 (not to scale).}
  \label{cartoon}
\end{figure*}

We suggest that the SNR G\,343.1$-$2.3 is the result of an off-centred
cavity SN explosion. Fig.~\ref{cartoon} schematically depicts a
scenario for its origin.
A massive star (the progenitor of the SN) ends its evolution within a
mushroom-like cavity formed by the SN progenitor wind breaking out of
the parent molecular cloud and expanding into an inter-cloud medium of
much less density. The proper motion of the progenitor star results
in a considerable offset of the SN explosion site from the geometrical
centre of the semi-spherical cavity created inside the cloud; we
suggest that the SN exploded outside the cloud. The subsequent
interaction of the SN blast wave with the reprocessed ambient medium
determines the structure of the resulting SNR (e.g. Ciotti \& D'Ercole
\cite{cio89}, Chevalier \& Liang \cite{che89}, Franco et al.\ 
\cite{fra91}), which acquires a form reminiscent of the well-known SNR
VRO\,42.05.01 (\object{G\,166.0+4.3})\footnote{It is believed that the
  unusual appearance of VRO\,42.05.01 is due to the breaking out of
  the SN blast wave into a hot, low-density tunnel, whose origin is
  the result of an amalgamation of cavities created by one or more SN
  explosions or stellar winds (e.g. Pineault et al.\ \cite{pin87}).
  We suggest that an alternative explanation of the origin of SNRs of
  this type (another example is the SNR \object{G\,350.0$-$3.0}) is a
  SN explosion inside a (mushroom-like) wind-driven cavity created
  near the edge of a molecular cloud. Numerical simulations of this
  situation would be highly desirable.}. 

We speculate that the wind-blown cavity formed inside the cloud was
surrounded by a shell of mass less than some critical value (for
spherically-symmetric shells this value is about 50 times the mass of
the SN ejecta; e.g. Franco et al.\ \cite{fra91}), so that the SN blast
wave was able to overrun the shell to propagate further into the
unperturbed gas of the cloud, leaving behind the reaccelerated and
gradually broadening turbulent shell (Franco et al.\ \cite{fra91}).
We suggest that the bright arc discovered by McAdam et al.\ corresponds to the shocked former wind-driven shell
and that the diffuse radio emission seen by Frail et al.\ and Dodson
et al.\ comes from the
``half" of the SN blast wave propagating into the cloud (see
Fig.~\ref{cartoon}). These two components correspond to the
  bright arclike structure in the low resolution image of
  G\,343.1$-$2.3 (Fig.\ 2).
Under this scenario a more extended component of the SNR
should exist to the southeast of the known bright structure.
This would correspond to the ``half" of the SN blast wave
expanding in the inter-cloud medium.

\begin{figure*}
  \resizebox{17cm}{!}{\includegraphics{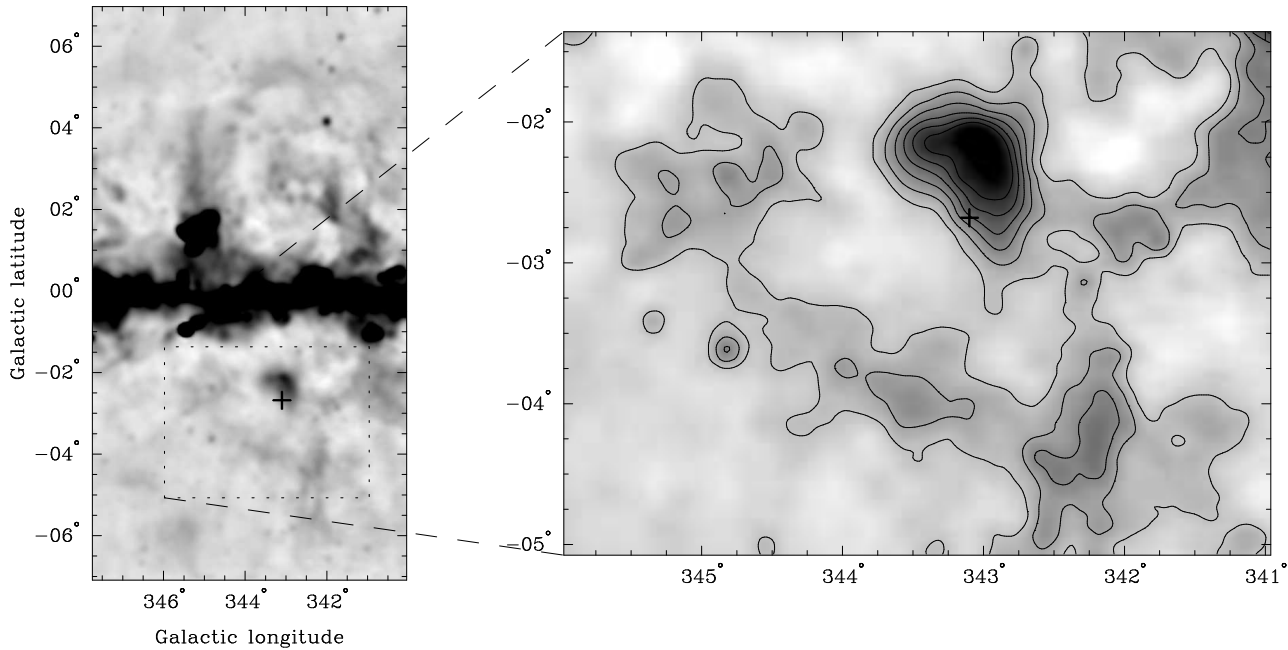}}
  \caption{2.4 GHz image of G\,343.1$-$2.3 (Duncan et al.\
    \cite{dun95}). A cross marks the position of the pulsar.}
  \label{plot}
\end{figure*}

We have found such a feature in the 2.4 GHz Parkes Survey of the
Galactic Plane (Duncan et al.\ \cite{dun95}; Fig.~\ref{plot}) -- an
extended arc of diffuse emission stretched from ($l=345$,
$b=-2.5$) to ($l=342$, $b=-4.5$).
Its location in the ``proper" place and its symmetry with respect to
the bright arc of G\,343.1$-$2.3 suggest that it could be physically
related to this SNR. We note that the ragged appearance of the
extended component
could result from the development of a Rayleigh-Taylor instability
caused by the impact of the SN blast wave with the wall of the
low-density cavity created in the inter-cloud medium by the SN
progenitor wind breaking out of the cloud (cf.\ Gvaramadze
\cite{gva99a}).  Note however that Duncan et al.\ 
consider this feature to be a part of a larger complex of filaments.
The association between the extended arc and the SNR G\,343.1$-$2.3
may be further investigated by detailed H{\tiny I} mapping in the
vicinity of the arc or by combining new continuum interferometric
observations with the existing single dish maps of the region.

\section{Reliability of the association between the pulsar PSR B\,1706$-$44 and
the SNR G\,343.1$-$2.3}

We now discuss the criteria for evaluating the reliability of
pulsar/SNR associations (Sect.~1) as applied to the system PSR
B\,1706$-$44/G\,343.1$-$2.3.  It is obvious that the first two
criteria should be fulfilled for any proposed pulsar/SNR association.
Application of the third and fifth ones for evaluating proposed
associations is not so straightforward, since they are based on the
assumption that the SN explosion site coincides with the geometrical
centre of the SNR. This assumption could be erroneous in the case of a
density-stratified interstellar medium (e.g. Lozinskaya \cite{loz92},
Frail et al.\ \cite{fra94b}) or in the case of a cavity SN explosion
(Gvaramadze \cite{gva02}). A mechanical application of these two
criteria could result in the rejection of genuine associations
(Gvaramadze \cite{gva02}). The fourth criterion should be considered
for those claimed associations where the pulsar is located not far (at
least in projection) from the SNR's shell. To some extent this
situation takes place in the case considered in this paper -- the
pulsar is superposed on the bright radio arc.

\subsection{The distance estimates}

The most reliable distance estimate for PSR B\,1706$-$44 was derived
by Koribalski et al.\ (\cite{kor95}) using a kinematic method. This
estimate ranges from $2.4\pm 0.6$ to $3.2\pm 0.4$ kpc, in quite
reasonable agreement with the dispersion measure distance of $1.8\pm
0.5$ kpc derived from the Taylor \& Cordes' (\cite{tay93}) model for
the Galactic electron density distribution, given the uncertainties
inherent in both methods. The kinematic distance measured by
Koribalski et al.\ also agrees with the distance estimates for 
G\,343.1$-$2.3 obtained by McAdam et al.\ (\cite{mca93}) and Frail et
al.\ (\cite{fra94a}). But the latter estimates are very uncertain,
since they are based on the highly controversial empirical
relationship between the observed surface brightness of SNRs and their
linear diameters (see e.g. Green \cite{gre91}; but see also Huang et
al.\ \cite{hua94}). These estimates would be even less certain if our
scenario for the origin of the SNR is correct. In what follows we
assume a distance to the pulsar (and the SNR) of $(1.8+2.4)/2=2.1$
kpc.

\subsection{The age estimates}

The characteristic age of PSR B\,1706$-$44 is $\tau =P/(n-1)\dot{P}$,
where $P$ is the spin period of the pulsar, $\dot{P}$ is the period
derivative, and $n$ is the braking index. For $P=0.102$ s, $\dot{P}
=9.3\cdot 10^{-14} \, {\rm s}\,{\rm s}^{-1}$ (Johnston et al.\ 
\cite{joh92}), and assuming that $n=3$, one has $\tau \simeq 17\,500$
yr.  This age can be compared with the age estimate for G\,343.1$-$2.3
of 5\,000--6\,000 yr, derived (McAdam et al. 1993, Nicastro et
al. 1996) on the basis of the
Sedov-Taylor solution (i.e.\ from the diameter-age relationship, e.g.
Clark \& Caswell \cite{cla76}). But the diameter-age relationship
cannot be applied to SNRs originating from cavity SN explosions.  In
what follows we assume that the true age of the pulsar is equal to
$\tau$ and that the SNR is as old as the pulsar. We realize, however,
that the system could be younger (if the pulsar was born with a spin
period close to the current one; in this case $\tau$ overestimates the
true age of the pulsar) or older (see Sect.\ 3.3.1), but the actual age
of the system is not fundamental to the results presented in this
paper.

\subsection{The pulsar velocity}

\subsubsection{The implied velocity}

The implied pulsar transverse velocity, i.e. the velocity inferred
from the angular displacement of PSR B\,1706$-$44 from the geometrical
centre of the bright arc, about $20'$ north of the pulsar, is
\begin{displaymath}
V_{\rm imp} \simeq 700 \theta _{20} D_{2.1} \tau _{17.5} ^{-1} \,
{\rm km}\,{\rm s}^{-1} \,\, ,
\end{displaymath}
where $\theta _{20}$ is the angular displacement in units of $20\arcmin$,
$D_{2.1}$ is the distance to the pulsar in units of 2.1 kpc, and $\tau
_{17.5}$ is the characteristic age of the pulsar in units of 17.5 kyr.
In principle this high transverse velocity is not
impossible. It could be even higher if the true age of the
  pulsar is less than $\tau$ (see Sect.\ 3.2).
However it can be reduced if one
assumes that the true age of the pulsar is much larger than the
characteristic one [e.g.\ due to the low value of the pulsar braking
index (e.g.\ Camilo \cite{cam96}) or due to a secular (e.g.
Blandford \& Romani \cite{bla88}) or short-term (e.g. Gvaramadze
\cite{gva99b}, \cite{gva01a}) increase of the braking
torque]. The implied velocity can also be reduced if the true SN
explosion site is offset from the geometrical centre of the SNR. Such
offsets naturally arise if the SN exploded in a density-stratified
medium (e.g. Lozinskaya \cite{loz92}, Frail et al.\ \cite{fra94b}) or
inside a cavity created by the wind of the {\it moving} SN progenitor
star (Gvaramadze \cite{gva02}). We favour the last possibility
(Sect. 2.2) and suggest that the pulsar transverse velocity could be
less than the implied one (see Sect. 3.3.2).

The implied transverse velocity should be compared with the measured
one. Nicastro et al.\ (\cite{nic96}) derived the pulsar velocity from the 
scintillation measurements and found that it is anomalously low, about 
twenty times less than $V_{\rm imp}$. They used this inconsistency to 
suggest that the pulsar did not originate from the
apparent centre of SNR, and that the pulsar and SNR are not
associated.  We agree with their first suggestion (see Sect.~2.2) and
therefore believe that the implied velocity can be reduced.  On the
other hand we have found (Sect.~3.3.2) that if the turbulent material
of the reaccelerated former wind-driven shell (the bright arc of SNR
G\,343.1$-$2.3) is responsible for nearly all the scattering of PSR
B\,1706$-$44, then the pulsar transverse velocity can be as large as
the transverse velocity of the portion of the arc projected on the
pulsar. If so, one can show that the pulsar (transverse) velocity
should indeed be less than $V_{\rm imp}$, though it can be much larger
than that calculated by Nicastro et al.\ (\cite{nic96}).

\subsubsection{The scintillation velocity}

It is known that pulsar velocities derived from scintillation
measurements show good correlation with proper-motion--derived ones
(e.g.~Gupta \cite{gup95}). However,
scintillation velocities are derived under several assumptions, some
of which are not necessarily suitable for individual pulsars. It is
usually assumed that the scattering material is homogeneously
distributed along the line of sight and that its transverse velocity
is negligible. These assumptions  could be erroneous if a
considerable fraction of the scintillations is due to a
localized region of enhanced scattering existing along the line of
sight in addition to the distributed scattering medium and if the
transverse velocity of this region is nonzero (e.g.\  Cordes \& Rickett
\cite{cor98}). In this case the scintillation velocity of the pulsar
is not equal to the proper motion velocity.  In the following we
assume that the turbulent material associated with the bright arc of
the SNR G\,343.1$-$2.3 is the main scatterer of PSR B\,1706$-$44.

The scintillation velocity for an asymmetrically placed thin
scattering screen is (e.g.~Gupta \cite{gup95}):
\begin{equation}
V_{\rm iss} = 3.85\times 10^4 \,{(\nu _{{\rm d},{\rm MHz}}
D_{\rm kpc} x)^{1/2} \over f_{\rm GHz}
t_{\rm d}} \, {\rm km}\,{\rm s}^{-1} \, ,
\label{v_iss_gupta95}
\end{equation}
where $\nu _{{\rm d},{\rm MHz}}$ and $t_{\rm d}$ are the scintillation
bandwidth and the time-scale measured respectively in MHz and seconds,
$D_{\rm kpc}$ is the distance from observer to pulsar in kpc,
$x=D_{\rm o} /D_{\rm p}$, $D_{\rm o}$ and $D_{\rm p}$ are the
distances from observer to screen and from screen to pulsar, and
$f_{\rm GHz}$ is the frequency of observation in units of GHz.  Note
that the numerical coefficient in Eq.\ (\ref{v_iss_gupta95}) is about three times larger
than that used by Nicastro et al.\ (\cite{nic96}; see, however,
Johnston et al.\ \cite{joh98}).  With $\nu_{{\rm d},{\rm MHz}} =
15$, $t_{\rm d} = 2287$, $f_{\rm GHz} =1.52$, and assuming that $x=1$
(Nicastro et al.\ \cite{nic96}), one has for PSR B\,1706$-$44 that
$V_{\rm iss} = 62D_{2.1} ^{1/2} \, {\rm km}\,{\rm s}^{-1}$. Although
this value is few times larger than that calculated by Nicastro et
al.~(\cite{nic96}), it is still at the lower limit of
the pulsar velocity distribution (e.g. Blaauw \& Ramachandran
\cite{bla98}, Lorimer et al.\ \cite{lor97}).  However, the pulsar
velocity estimate can be further increased if one takes into
account the transverse velocity of the scattering screen.

In the observer's reference frame the scintillation velocity is
connected with the pulsar (transverse or proper motion) velocity $V_{\rm p}$ by the
following relationship (Gvaramadze \cite{gva01b}; cf. Gupta et
al.~\cite{gup94}, Cordes \& Rickett \cite{cor98}):
\begin{eqnarray}
\label{v_iss_gva01}
V_{\rm iss} &=& \Big[x^2 V_{\rm p} ^2 -2x(1+x)V_{\rm p} V_{{\rm scr},\parallel}
 \nonumber \\
      &+&(1+x)^2 V_{{\rm scr},\parallel} ^2 + (1+x)^2 V_{{\rm scr},\perp} ^2 \Big]^{1/2} \, ,
\end{eqnarray}
where $V_{{\rm scr},\parallel}$ and $V_{{\rm scr},\perp}$ are the
components of the transverse velocity of the screen, correspondingly,
parallel and perpendicular to the vector of the pulsar proper motion
velocity. In (\ref{v_iss_gva01}) we neglected for simplicity contributions from the
differential Galactic rotation and the Earth's orbital motion around
the Sun. If $V_{\rm scr} =0$ [as assumed by Nicastro et al.
(1996)], one has $V_{\rm p} = V_{\rm iss} /x \simeq 62 x^{-1/2}
D_{2.1}^{1/2} \, {\rm km}\,{\rm s}^{-1} \simeq 0$ (note that $x\sim
100$).  But, if $V_{\rm scr} \neq 0$, one can solve Eq.\ (\ref{v_iss_gva01}) for
$V_{\rm p}$:
\begin{displaymath}
V_{\rm p} \simeq V_{{\rm scr},\parallel} \pm (V_{\rm iss} ^2 /x^2 -
V_{{\rm
scr},\perp} ^2 )^{1/2} \, \, .
\end{displaymath}
This solution is physically meaningful only for $V_{{\rm scr},\perp} \leq
V_{\rm iss} /x \, (\simeq 0)$, i.e. if the pulsar moves in
the same direction and with nearly the same (transverse) velocity as does
the
part of the SNR responsible for the scattering of the pulsar:
$V_{\rm p} \simeq V_{{\rm scr},\parallel}$.

Although the existing observational data do not allow us to estimate
the expansion velocity of the bright arc, we can constrain it by
setting an upper limit on the expansion velocity of the ``half" of the
SN blast wave propagating inside the cloud. This constraint can be
derived from the non-detection of the soft (0.1--2.4 keV) X-ray
emission from the SNR.  For the column density towards the SNR of
$\simeq 2-5\cdot 10^{21} \, {\rm cm}^{-2}$ (Becker et al.\ \cite{bec95}), the
interstellar medium transmits essentially no X-ray emission with
energies below 0.3 keV (e.g. Gorenstein \& Tucker \cite{gor76}),
therefore the expansion velocity of the blast wave is less than
$\simeq 500 \, {\rm km}\,{\rm s}^{-1}$. On the other hand, it is
obvious that the former wind-driven shell lagging behind the blast
wave (whose angular extent is about two times
larger) expands at least two times more slowly. These arguments show that
if the bright arc is the main scatterer of the pulsar's radio
emission, then the pulsar transverse velocity should indeed be less
than $V_{\rm imp}$, though it could be as large as $200 \, {\rm
  km}\,{\rm s}^{-1}$.

\subsection{Interaction between the pulsar and the SNR}

We mentioned above that despite the apparent proximity of PSR
B\,1706$-$44 to the bright arc of the SNR G\,343.1$-$2.3 there are no
morphological signatures of interaction between these two objects.
This has given some authors a basis to question their association
(Frail et al.\ \cite{fra94a}; see also Nicastro et al.\ 
\cite{nic96}). But this inconsistency can be easily removed if the
SN exploded within a mushroom-like wind-driven cavity (Sect.~2.2).

\subsection{Orientation of the pulsar proper motion vector}

The proper motion vectors of neutron stars born in off-centred cavity
SN explosions can be oriented arbitrarily with respect to the
geometric centres of the associated SNRs (Gvaramadze \cite{gva02});
they can even be directed towards the geometric centres of the
SNRs!\footnote{Perhaps exactly this situation takes place in the case
  of the pulsar \object{PSR B\,0656$+$14}, which is moving
  approximately towards the centre of the SNR \object{Monogem Ring}
  (Thompson \& C{\'o}rdova \cite{tho94}).} In Sect.~3.3 we showed that if
the scintillations of PSR B\,1706$-$44 are due mostly to the
scattering in the turbulent material associated with the bright arc of
SNR G\,343.1$-$2.3, then the pulsar proper motion should be parallel
to the expansion velocity of this material, i.e.~the pulsar should
move from the northeast to the southwest; this should be tested
observationally.

\subsection{On statistical studies of pulsar/SNR associations}

Now we discuss the statistical argument against the association
between PSR B\,1706$-$44 and the SNR G\,343.1$-$2.3 mentioned by
Nicastro et al.\  (\cite{nic96}). This argument is based on the
result of a statistical study of pulsar/SNR associations by Gaensler \&
Johnston (\cite{gae95}), which suggests that young ($<25\,000$ yr)
pulsars cannot overrun their parent SNR shells (we recall that the
spin-down age of PSR B\,1706$-$44 is $\simeq 17\,500$ yr).  Although
it is now clear that PSR B\,1706$-$44 is located (at least in
projection) well within the SNR G\,343.1$-$2.3 (Dodson et al.\
\cite{dod01}), it should be mentioned that Gaensler \& Johnston
(\cite{gae95}) did not consider two very important
effects: modification of the ambient medium by the ionizing emission
and stellar wind of massive stars (the progenitors of most SNe), and
the proper motion of SN progenitor stars. Taking into account these
two effects allows it to be shown that even a young pulsar moving with
a moderate velocity ($\simeq 200\, {\rm km}\,{\rm s}^{-1}$) is
able to escape the SNR's shell, provided that it was born not far from
the edge of the wind-driven bubble (Gvaramadze
\cite{gva02}).  Alternatively, the apparent location of a pulsar
on the edge of a SNR's shell can be due simply to the effect of
projection in non-spherically-symmetric SNRs (see Fig.~\ref{cartoon}).

\subsection{On the wind nebula around PSR B\,1706$-$44}

The large angular extent of the pulsar wind nebula was used by Frail
et al.\ (\cite{fra94a}) to consider an association between the
pulsar and the SNR unlikely. Assuming that the pulsar moves (with a
velocity of $670 \, {\rm km}\,{\rm s}^{-1}$) through the interstellar
medium (of number density $1\, {\rm cm}^{-3}$) and that the pulsar
wind nebula is confined by the ram pressure of the ambient medium,
they found that the characteristic radius of the nebula, $r_{\rm PWN}$,
should be about two orders of magnitude less than observed. 
Based on this
argument Frail et al.\ suggested that G\,343.1$-$2.3 could be a
background object, while PSR B\,1706$-$44 could be a low-velocity
pulsar. We agree with their last suggestion, but propose that the
pulsar wind nebula is instead confined by the interaction with
the hot, tenuous material which fills the interior of the SNR:
\begin{equation}
{|\dot{E}| \over 4\pi cr_{\rm PWN} ^2} \simeq n(2kT + \mu m_{\rm H}  
v_{\rm p} ^2 ) \, ,
\end{equation}
where $|\dot{E}| =3.4\times 10^{36} \, {\rm ergs} \, {\rm s}^{-1}$ is
the pulsar spin-down luminosity, $c$ is the speed of light, $n$ and
$T$ are, respectively, the number density and the temperature inside
the SNR, $k$ is the Boltzmann constant, $\mu =1.3$ is the mean
molecular weight, $m_{\rm H}$ is the mass of a hydrogen atom, and
$v_{\rm p}$ is the full (i.e.\ three-dimensional) pulsar velocity. For
$r_{\rm PWN} \simeq 0.9 \, D_{2.1} ^{-1}$ pc and $v_{\rm p} \simeq 200
\, {\rm km} \, {\rm s}^{-1}$, and assuming that $T\simeq 10^7$ K, one
has from Eq. (3) a quite reasonable estimate for the number density,
$n = 3.2\times10^{-4} \, {\rm cm}^{-3}$.  Note that this estimate
could be somewhat altered should the pulsar velocity have a
significant radial component (a case not constrained by the
considerations of this paper).  For example, if $v_{\rm p} = 500 \,
{\rm km}\, {\rm s}^{-1}$, then one has $n =1.4\times10^{-4} \, {\rm
  cm}^{-3}$ (i.e.\ also a reasonable value).

\section{Conclusion}

We have analyzed the available observational data on the pulsar PSR
B\,1706$-$44 and the SNR G\,343.1$-$2.3 and suggested that these
objects could be the remnants of a SN which exploded within a
mushroom-like cavity created by the SN progenitor wind breaking out of
the parent molecular cloud.  This accounts for the disparity between
the measured and implied velocities of the pulsar. Our suggestion
implies that in addition to the known bright ``half" of the SNR
G\,343.1$-$2.3 there should exist a more extended and weaker
component, so that the actual shape of G\,343.1$-$2.3 is similar to
that of the well-known SNR VRO\,42.05.01. We have found such a
component in archival radio data. Further observations, such as those
discussed in Sect.\ 2.2, would be useful to confirm or reject the
association between this component and the SNR G\,343.1-2.3.

\begin{acknowledgements}

  We are grateful to R.\,Dodson for providing his manuscript in
  advance of publication, and to the referee (J.\ Cordes) for several
  useful suggestions. V.V.\,Gvaramadze is supported in part by the
  Deutscher Akademischer Austausch Dienst (DAAD).

\end{acknowledgements}

\end{document}